\documentclass[12pt]{iopart}%
\usepackage{graphicx}
\usepackage{tabularx}
\usepackage{dcolumn}
\usepackage{tensor}
\usepackage{placeins}
\usepackage{bm}
\usepackage{amsfonts}
\usepackage{amssymb}
\usepackage{amsopn}
\expandafter\let\csname equation*\endcsname=\relax 
\expandafter\let\csname endequation*\endcsname=\relax 
\usepackage{amsmath}
\usepackage{float}%
\providecommand{\U}[1]{\protect\rule{.1in}{.1in}}
\begin{document}
\title{On Hagedorn wavepackets associated with different Gaussians}
\author{Ji\v{r}\'{\i} J. L. Van\'{\i}\v{c}ek, Zhan Tong Zhang}
\address{Laboratory of Theoretical Physical Chemistry, Institut des Sciences et
Ing\'enierie Chimiques, Ecole Polytechnique F\'ed\'erale de Lausanne (EPFL),
CH-1015 Lausanne, Switzerland}
\eads{\mailto{jiri.vanicek@epfl.ch}, \mailto{zhan.zhang@epfl.ch}}
\date{\today}

\begin{abstract}
Hagedorn functions are carefully constructed generalizations of Hermite
functions to the setting of many-dimensional squeezed and coupled harmonic
systems. Wavepackets formed by superpositions of Hagedorn functions have been
successfully used to solve the time-dependent Schr\"{o}dinger equation exactly
in harmonic systems and variationally in anharmonic systems. To evaluate
typical observables, such as position or kinetic energy, it is sufficient to
consider orthonormal Hagedorn functions with a single Gaussian center. Instead, we derive various relations between Hagedorn bases associated with
different Gaussians, including their overlaps, which are necessary for
evaluating quantities nonlocal in time, such as the time correlation functions
needed for computing spectra. First, we use the Bogoliubov transformation to
obtain the commutation relations between the ladder operators associated with
different Gaussians. Then, instead of using numerical quadrature, we employ
these commutation relations to derive exact recurrence relations for the
overlap integrals between Hagedorn functions with different Gaussian centers.
Finally, we present numerical experiments that demonstrate the accuracy and
efficiency of our algebraic method as well as its suitability for treating
problems in spectroscopy and chemical dynamics.

\end{abstract}
\maketitle

\graphicspath{
{./Figures/}{C:/Users/Jiri/Dropbox/Papers/Chemistry_papers/2024/algebraic_overlap/Figures/}}

\section{Introduction}

Heller~\cite{Heller:1975,Heller:1981a} and Hagedorn~\cite{Hagedorn:1981} were
among the first to use semiclassical Gaussian wavepackets to approximate the
solutions of the nuclear time-dependent Schr\"{o}dinger equation (TDSE). They
were motivated by the fact that these wavepackets are, in fact, exact
solutions in multidimensional harmonic systems. Although many modern dynamical
methods employ multiple
Gaussians~\cite{Herman_Kluk:1984,Ben-Nun_Martinez:2000,Worth_Burghardt:2004,Beck_Meyer:2000,Miller:2001,Werther_Grossmann:2020},
single Gaussian wavepacket dynamics~\cite{Lasser_Lubich:2020,
Vanicek:2023}, such as the
thawed~\cite{Heller:1975,Heller:1981a,book_Heller:2018} and variational
\cite{Heller:1976,Hellsing_Metiu:1985,Arickx_VanLeuven:1986,Coalson_Karplus:1990,book_Lubich:2008} Gaussian
approximations, have seen a resurgence in their applications in chemical
dynamics and vibronic
spectroscopy~\cite{Frantsuzov_Mandelshtam:2004,Grossmann:2006,Deckman_Mandelshtam:2010,Cartarius_Pollak:2011,Wehrle_Vanicek:2014,Gottwald_Kuehn:2019,Begusic_Vanicek:2019,Golubev_Vanicek:2020,Begusic_Vanicek:2020,Begusic_Vanicek:2022,Scheidegger_Golubev:2022,Fereidani_Vanicek:2023,Fereidani_Vanicek:2024,Poulsen_Nyman:2024,Gherib_Genin:2024,Ryabinkin_Genin:2024}.
Although single-Gaussian methods cannot capture wavepacket splitting
and are, in general, limited to short-time dynamics in weakly anharmonic
systems, they provide substantial improvements over global harmonic models.

To describe the distortion of a Gaussian during propagation and, more
broadly, to propagate non-Gaussian wavepackets, Hagedorn devised an elegant
orthonormal basis, which generalizes the Hermite basis for a simple harmonic
oscillator, is guided by a semiclassical Gaussian, and permits the expansion of an
arbitrary wavepacket. In the case of one-dimensional quantum harmonic
oscillators, the solution of the time-independent Schr\"{o}dinger equation
yields equally separated energy eigenvalues and eigenfunctions in the form of
Hermite polynomials multiplied by a Gaussian function.
The same solution can be obtained using an algebraic approach of raising and
lowering ``ladder'' operators introduced by Dirac.~\cite{book_Dirac:1947,book_Tannor:2007}.

Hagedorn adopted an analogous approach and introduced a set of raising and
lowering operators that can be applied to a general multidimensional
Gaussian~\cite{Hagedorn:1981,Hagedorn:1985,Hagedorn:1998}. Starting from a
Gaussian wavepacket, these operators generate a complete orthonormal basis
consisting of so-called Hagedorn functions, which are products of specific
polynomials with the original Gaussian. Remarkably, similar to the Gaussian
wavepacket, each Hagedorn function is also an exact solution to the TDSE with
a harmonic potential. Superpositions of Hagedorn functions, called Hagedorn
wavepackets, can be used to approximate the solutions to the TDSE in arbitrary
orders of $\hbar$~\cite{Hagedorn:1998,Lasser_Lubich:2020}. While the Gaussian
center is propagated in the same way as in the Gaussian wavepacket dynamics,
the coefficients of the basis functions remain constant in harmonic potentials
and can be propagated variationally in non-quadratic
potentials~\cite{book_Lubich:2008,Lasser_Lubich:2020}. Consequently, Hagedorn
wavepackets are much more suitable for treating weakly anharmonic many-dimensional problems
than are computationally expensive grid-based numerical
methods~\cite{Faou_Lubich:2009,Gradinaru_Hagedorn:2010,Zhou:2014,Gradinaru_Rietmann:2021}.
They have attracted significant attention in the mathematical literature~\cite{Faou_Lubich:2009,Hagedorn:2013,Ohsawa_Leok:2013,Gradinaru_Hagedorn:2014,Lasser_Troppmann:2014,Li_Xiao:2014,Hagedorn:2015,Ohsawa:2015,Punosevac_Robinson:2016,Dietert_Troppmann:2017,Bourquin:2017,Hagedorn_Lasser:2017,Lasser_Troppmann:2018,Ohsawa:2018,Punosevac_Robinson:2019,Ohsawa:2019,Blanes_Gradinaru:2020,Arnaiz:2022,Miao_Zhou:2023}
with several applications in physics~\cite{Kargol:1999,Hagedorn_Joye:2000,Gradinaru_Joye:2010,Gradinaru_Joye:2010a,Kieri_Karlsson:2012,Bourquin_Hagedorn:2012,Zhou:2014,Gradinaru_Rietmann:2021,Gradinaru_Rietmann:2024}.
For detailed and mathematically rigorous reviews of the properties of Hagedorn
wavepackets, see~\cite{book_Lubich:2008,Lasser_Lubich:2020}.

The orthonormality of the Hagedorn basis avoids many numerical issues
encountered by methods, such as multi-trajectory Gaussian-basis techniques,
that employ nonorthogonal bases. In a Hagedorn wavepacket, it is
straightforward to evaluate the expectation values of observables local in time,
such as position, momentum, or kinetic energy. In contrast, the application of
Hagedorn wavepackets in spectroscopy has been limited because the spectrum
depends on the wavefunction at all times (up to the time that determines the
spectral resolution). Specifically, the spectrum is given by the Fourier
transform of the wavepacket autocorrelation function~\cite{book_Heller:2018},
the overlap between the initial and propagated wavepackets, which is
numerically difficult to evaluate because the initial and final Hagedorn
wavepackets are expanded in different, mutually nonorthogonal Hagedorn bases
associated with the initial and propagated Gaussians. Overlaps of highly
excited Hagedorn functions result in highly oscillatory integrals that are
difficult to evaluate numerically in high dimensions, and may even encounter
problems due to the finite precision of computers~\cite{Bourquin:2017}. Because
standard numerical methods, including Gauss--Hermite quadratures, are
insufficient, more sophisticated numerical algorithms, such as those based on
sparse grids have been proposed~\cite{Faou_Lubich:2009,Bourquin:2017}.

Here, we avoid numerical approaches altogether and instead derive an exact algebraic scheme for computing the overlap
between arbitrary Hagedorn functions or wavepackets with different Gaussian
centers. Although the integrals of multivariate Gaussians multiplied by an explicit
polynomial are known, no explicit form is currently available for the
polynomial prefactors of Hagedorn wavepackets. In a remarkable tour de force,
Lasser and Troppmann derived an analytical expression for the
Fourier--Bros--Iagolnitzer transform of any Hagedorn function, which is a
special case of the overlap of a Hagedorn function with a (spherical)
Gaussian~\cite{Lasser_Troppmann:2014}. In contrast, our exact expression is
only recursive but applies directly to more general situations where both
states are arbitrary Hagedorn functions. The main result of our study is
this overlap expression, which should find interesting applications in
spectroscopy. Yet, we also obtain on the way various other useful relations between
Hagedorn operators and functions associated with two different Gaussians.

The remainder of this paper is organized as follows. In Sec.~II, we review
Hagedorn operators, functions, and wavepackets with a single Gaussian center.
In Sec.~III, we describe the Bogoliubov transformation and commutation
relations between the ladder operators associated with different Gaussians and
relate the results to the canonical symplectic structure on phase space. In
Sec.~IV, we derive and solve a system of linear equations for the overlaps of
higher-order Hagedorn functions with two different Gaussian centers in terms
of the overlaps of lower-order Hagedorn functions. This solution provides a
recursive algorithm for the overlap between two arbitrary Hagedorn functions, because the
formula for the overlap of two Gaussian wavepackets (i.e., zeroth-order
Hagedorn basis functions) is well-known. Section~V contains numerical
experiments that demonstrate that the recursive expression for the overlap is
accurate, efficient, robust, and applicable to higher dimensional problems in
chemical dynamics.
For a three-dimensional harmonic system, we also compare the autocorrelation
function obtained with our algorithm from the propagated Hagedorn wavepacket
to the autocorrelation function computed numerically from the exact quantum
split-operator propagation.

\section{Hagedorn wavepackets associated with a single Gaussian}

We begin by reviewing the construction of Hagedorn functions and
wavepackets with a single Gaussian center and by defining notation that will
be useful in later sections.

\subsection{Canonical symplectic structure on phase space}

Let $I_{D}$ be the $D$-dimensional identity matrix and%
\begin{equation}
J=%
\begin{pmatrix}
0 & I_{D}\\
-I_{D} & 0
\end{pmatrix}
\label{eq:J_mat}%
\end{equation}
the $2D\times2D$-dimensional standard symplectic matrix. $J$ defines a
canonical symplectic structure $\omega$ on phase space, i.e., a nondegenerate
skew-symmetric bilinear form, which for any $2D$-dimensional phase-space
vectors $
z:=\binom{q}{p}$ and $z^{\prime}:=\binom{q^{\prime}}{p^{\prime}}$
gives the real number%
\begin{equation}
\omega(z,z^{\prime})=z^{T}\cdot J\cdot z^{\prime}. \label{eq:omega_for_z}%
\end{equation}
We use (and slightly abuse) the notation $\omega$ more generally, so that
for any $2D\times D_{1}$ complex matrix $X$ and $2D\times D_{2}$ complex matrix $Y$, the
expression%
\begin{equation}
\omega(X,Y):=X^{T}\cdot J\cdot Y \label{eq:omega_for_any_X_Y}%
\end{equation}
yields a $D_{1}\times D_{2}$ complex matrix.

\subsection{Gaussian wavepacket in Hagedorn parametrization}

In Hagedorn's
parametrization~\cite{Hagedorn:1998,book_Lubich:2008,Vanicek:2023}, a
normalized complex-valued $D$-dimensional Gaussian wavepacket is written as
\begin{equation}
g[\Lambda_{t}](q)=\frac{1}{(\pi\hbar)^{D/4}\sqrt{\det(Q_{t})}}%
\exp\left[  \frac{i}{\hbar}\left(  \frac{1}{2}x^{T}\cdot P_{t}\cdot Q_{t}%
^{-1}\cdot x+p_{t}^{T}\cdot x+S_{t}\right)  \right]  ,\label{eq:tga}%
\end{equation}
with the shifted position $x:=q-q_{t}$ and a set of time-dependent parameters
$\Lambda_{t}=(q_{t},p_{t},Q_{t},P_{t},S_{t})$, where $q_{t}$ and $p_{t}$
represent the position and momentum of the center of the wavepacket. Whereas
Heller's parametrization uses a complex, symmetric $D$-dimensional width
matrix $C_{t}$ with a positive definite imaginary part and a complex phase
factor $\gamma_{t}$, here the width matrix $C_{t}=P_{t}\cdot Q_{t}^{-1}$ is
factorized into two complex $D$-dimensional matrices and the real phase factor
$S_{t}$ is related to the classical action. Hagedorn's parametrization offers
classical-like equations of motion for the components related to the width of
the Gaussian~\cite{Hagedorn:1998,Lasser_Lubich:2020,Vanicek:2023} and
facilitates the algebraic construction of higher-order Hagedorn functions,
which is described in Sections \ref{ss:ladder_operators} and \ref{ss:hgd_func}.

The complex matrices $Q_{t}$ and $P_{t}$ are related to the position and momentum covariances~\cite{Vanicek:2023}
and must satisfy the conditions~\cite{book_Lubich:2008,Lasser_Lubich:2020}%
\begin{align}
Q_{t}^{T}\cdot P_{t}-P_{t}^{T}\cdot Q_{t} &  =0,\label{eq:Q_P_sympl_1}\\
Q_{t}^{\dag}\cdot P_{t}-P_{t}^{\dag}\cdot Q_{t} &  =2iI_{D}.\label{eq:Q_P_sympl_2}%
\end{align}
which are equivalent to requiring that the real $2D\times2D$
matrix%
\begin{equation}
Y:=%
\begin{pmatrix}
\operatorname{Re}Q_{t} & \operatorname{Im}Q_{t}\\
\operatorname{Re}P_{t} & \operatorname{Im}P_{t}%
\end{pmatrix}
\end{equation}
be symplectic, i.e., $Y^{T}\cdot J\cdot Y=J$.
In addition~\cite{Lasser_Lubich:2020},
$
\operatorname{Im}C=(Q_{t}\cdot Q_{t}^{\dag})^{-1},\label{eq:ImA_from_Q}%
$
and both $Q_{t}\cdot Q_{t}^{\dag}$ and $P_{t}\cdot P_{t}^{\dag}$ are symmetric
matrices%
\begin{align}
(Q_{t}\cdot Q_{t}^{\dag})^{T} &  =Q_{t}\cdot Q_{t}^{\dag},\label{eq:QQdag_sym}%
\\
(P_{t}\cdot P_{t}^{\dag})^{T} &  =P_{t}\cdot P_{t}^{\dag}.\label{eq:PPdag_sym}%
\end{align}
Symmetry of $Q_{t}\cdot Q_{t}^{\dag}$ and $P_{t}\cdot P_{t}^{\dag}$ is
equivalently expressed by the relations%
\begin{equation}
\bar{Q}_{t}\cdot Q_{t}^{T}=Q_{t}\cdot Q_{t}^{\dag}\text{ \ \ and \ \ }\bar
{P}_{t}\cdot P_{t}^{T}=P_{t}\cdot P_{t}^{\dag}.\label{eq:QQdag_or_PPdag_T}%
\end{equation}
Every complex symmetric matrix $C_{t}$ with a positive definite imaginary part
can be factorized into two matrices that satisfy symplecticity conditions
(\ref{eq:Q_P_sympl_1}) and (\ref{eq:Q_P_sympl_2})~\cite{book_Lubich:2008}.
However, this factorization is not unique. For convenience, given a
Gaussian initial state with a known width matrix $C_{0}$ (e.g., from
electronic structure calculations), we choose $Q_{0}=(\operatorname{Im}%
C_{0})^{-1/2}$ and $P_{0}=C_{0}\cdot Q_{0}$.

From a mathematical point of view, Gaussian wavepackets $g[\Lambda_{t}](q)$ form a finite-dimensional submanifold of the Hilbert space $L^{2}%
(\mathbb{R}^{D})$ of square-integrable functions on $\mathbb{R}^{D}$. For each Gaussian, its tangent vectors are precisely
all functions obtained from this Gaussian by multiplication by at most
quadratic polynomials. As a result, the Gaussian wavepacket $g[\Lambda_{t}](q)$ preserves its form at all times and exactly solves the
TDSE when the potential function is at most quadratic and the parameters
$(q_{t},p_{t},Q_{t},P_{t},S_{t})$ solve a classical-like system of ordinary
differential equations~\cite{book_Lubich:2008}. Moreover, the symplecticity relations
(\ref{eq:Q_P_sympl_1}) and (\ref{eq:Q_P_sympl_2}) remain satisfied at all
times~\cite{Lasser_Lubich:2020}. Remarkably, none of these properties of the
Gaussian wavepacket are lost even when the quadratic potential depends on
time~\cite{Lasser_Lubich:2020} or on the wavepacket itself~\cite{Vanicek:2023}.
For example, if an arbitrary potential is approximated with the local harmonic
approximation, one obtains Heller's celebrated thawed Gaussian
approximation~\cite{Heller:1975}, which has been applied to solve a wide range
of spectroscopic problems beyond global harmonic
models~\cite{Grossmann:2006,Wehrle_Vanicek:2014,Wehrle_Vanicek:2015,Begusic_Vanicek:2020a,Prlj_Vanicek:2020,Begusic_Vanicek:2021a,Kletnieks_Vanicek:2023}.

\subsection{Raising and lowering operators}
\label{ss:ladder_operators}

Generalizing Dirac's construction for a one-dimensional harmonic oscillator,
Hagedorn constructed an orthonormal basis of $L^{2}(%
\mathbb{R}
^{D})$ by applying certain raising operators to the Gaussian state
(\ref{eq:tga}). In the following, we suppress the time subscript $t$ on
all quantities except for parameters $q_{t}$ and $p_{t}$, where the subscript is
necessary for distinguishing the parameters $q_{t}$ and $p_{t}$ of the Gaussian from the
arguments $q$ and $p$ of the wavefunction in position or momentum representation.

To simplify notation, let us define the shifted position and momentum
operators%
\begin{equation}
\hat{x}:=\hat{q}-q_{t}\text{ \ \ and \ \ }\hat{\xi}:=\hat{p}-p_{t}%
\label{eq:shifted_q_p}%
\end{equation}
with zero expectation values ($\langle\hat{x}\rangle=\langle\hat{\xi}%
\rangle=0$) in the Gaussian wavepacket. Hagedorn introduced the lowering and
raising $D$-dimensional vector operators%
\begin{align}
A &  :=-\frac{i}{\sqrt{2\hbar}}\left(  P^{T}\cdot\hat{x}-Q^{T}\cdot\hat{\xi
}\right)  ,\label{eq:A}\\
A^{\dag} &  :=\frac{i}{\sqrt{2\hbar}}\left(  P^{\dag}\cdot\hat{x}-Q^{\dag
}\cdot\hat{\xi}\right)  .\label{eq:Adag}%
\end{align}
With $Q$ and $P$ satisfying the symplecticity relations
(\ref{eq:Q_P_sympl_1}) and (\ref{eq:Q_P_sympl_2}), the components of the two
operators enjoy the commutator relations
\begin{equation}
\lbrack A_{j},A_{k}^{\dagger}]=\delta_{jk}\text{ \ \ and \ \ }[A_{j}%
,A_{k}]=[A_{j}^{\dag},A_{k}^{\dag}]=0,\label{eq:comm_ladder}%
\end{equation}
for $j,k = 1,...,D$.
In one-dimensional cases, the two operators reduce to Dirac's well-known
ladder operators. The shifted position and momentum operators can be recovered
from the raising and lowering operators as%
\begin{align}
\hat{x} &  =\sqrt{\hbar/2}(\bar{Q}\cdot A+Q\cdot A^{\dag}%
),\label{eq:x_from_A_Adag}\\
\hat{\xi} &  =\sqrt{\hbar/2}(\bar{P}\cdot A+P\cdot A^{\dag}%
).\label{eq:xi_from_A_Adag}%
\end{align}

\subsection{Hagedorn functions}
\label{ss:hgd_func}

The zeroth-order Hagedorn function $\varphi_{0}:=g$ is defined to be the Gaussian
wavepacket in (\ref{eq:tga}). Other Hagedorn functions $\varphi_{K}$ associated
with a Gaussian $\varphi_{0}$ are parametrized with a multi-index
$K=(K_{1},\dots,K_{D})\in\mathbb{N}_{0}^{D}$ and recursively generated by
applying the raising operator, $\varphi_{K+\langle j\rangle}=\frac{1}%
{\sqrt{K_{j}+1}}A_{j}^{\dagger}\varphi_{K}$, where $\langle j\rangle
=(0,\dots,0,1,0,\dots,0)$ denotes the $D$-dimensional unit vector with nonzero
$j$th component~\cite{book_Lubich:2008}. Indeed, both the raising and lowering
operators owe their names to the way they act on the Hagedorn functions:%
\begin{align}
A_{j}\varphi_{K}  &  =\sqrt{K_{j}}\varphi_{K-\left\langle j\right\rangle
},\label{eq:A_phiK}\\
A_{j}^{\dag}\varphi_{K}  &  =\sqrt{K_{j}+1}\varphi_{K+\left\langle
j\right\rangle }. \label{eq:Adag_phiK}%
\end{align}
In other words, lowering operator $A_{j}$ reduces the $j$th component of
the multi-index $K$ by $1$, whereas raising operator $A_{j}^{\dag}$
increases the $j$th component of $K$ by $1$. Owing to the commutation relations
(\ref{eq:comm_ladder}), different components of the $A^{\dagger}$ and $A$ vectors act
independently to increase and decrease $K$ in different degrees of freedom.

If expressed in position representation, Hagedorn functions take the form of a
Gaussian multiplied by a polynomial of degree $|K|=K_{1}+\dots+K_{D}$. These
polynomial prefactors (\textquotedblleft Hagedorn
polynomials\textquotedblright) have been studied in
detail~\cite{Dietert_Troppmann:2017,Hagedorn:2015,Ohsawa:2019}; however, we do
not have an explicit closed-form expression for them. They are connected to
the Hermite polynomials through squeezing and rotation
operators~\cite{Ohsawa:2019}, but for $D>1$ they are not, in general, simple
tensor products of one-dimensional Hermite polynomials~\cite{Hagedorn:1985}.

A special case occurs when the matrix product $Q^{-1}\cdot\overline{Q%
}$ is diagonal. The polynomial prefactor $\mathrm{Pol}_{K}$ of $\varphi_{K}$
can then be expressed as a direct product
\begin{equation}
\mathrm{Pol}_{K}(q)=\prod_{j=1}^{D}\lambda_{j}^{K_{j}/2}H_{K_{j}}\left(
\frac{q_{j}}{\sqrt{\lambda_{j}}}\right)  \label{eq:pol_direct}%
\end{equation}
of scaled Hermite polynomials $H_{K_{j}}$, where $\lambda_{j}$'s are the
eigenvalues of $Q^{-1}\cdot\overline{Q}=\operatorname{diag}%
(\lambda_{1},\dots,\lambda_{D})$. In an appropriate coordinate system,
Hagedorn functions can therefore easily represent the vibrational
eigenfunctions of a harmonic oscillator. Consequently, we sometimes refer
to $K$ as the \textquotedblleft excitation\textquotedblright\ of the Hagedorn
function $\varphi_{K}$ and we shall do so even when the condition for
(\ref{eq:pol_direct}) is not satisfied.

\subsection{Hagedorn wavepackets}

For any $\Lambda$, the Hagedorn functions form a complete orthonormal basis in
$L^{2}(\mathbb{R}^{D})$; therefore, we can expand an arbitrary solution
$\psi(t)$ of the TDSE as their superposition, called the Hagedorn wavepacket
$h(\mathbf{c},\Lambda)$:
\begin{equation}
\psi(t)\equiv h(\mathbf{c}_{t},\Lambda_{t}):=\sum_{K}c_{K}(t)\varphi
_{K}[\Lambda_{t}], \label{eq:HWP}%
\end{equation}
where $c_{K}(t)$ are complex-valued coefficients, and the basis functions
$\varphi_{K}$ are time-dependent only via the Gaussian parameters $\Lambda
_{t}$ defining $\varphi_{0}$ and the ladder operators. In practice, the
infinite-dimensional basis must be truncated to a finite basis by constraining
the multi-index $K$ to be only in a finite subset $\mathcal{K}%
\subset\mathbb{N}_{0}^{D}$~\cite{Faou_Lubich:2009}.

A beautiful property of the Hagedorn wavepackets is that if one employs the
global or
local harmonic approximation for the potential, the coefficients $c_{K}(t)$ do
not change with time and one only needs to propagate the Gaussian
parameters---in exactly the same classical-like way as in the thawed Gaussian
approximation. Alternatively, the coefficients $c_{K}$ can be propagated with
the variational principle to include the effects from the potential beyond the
local harmonic potential~\cite{Faou_Lubich:2009}.

Let us introduce a more succinct notation $K(\Lambda)$ for the Hagedorn
function $\varphi_{K}(\Lambda)$ and let us even suppress the argument
$\Lambda$ if all Hagedorn functions have the same Gaussian center. As such
Hagedorn functions are orthonormal,%
\begin{equation}
\langle J,K\rangle=\delta_{JK},\label{eq:HF_ON}%
\end{equation}
the scalar product of the Hagedorn wavepackets $\psi\equiv h(\mathbf{c},\Lambda)$
and $\psi^{\prime}\equiv h(\mathbf{c}^{\prime},\Lambda)$ associated with the
same Gaussian can be computed as%
\begin{equation}
\langle\psi,\psi^{\prime}\rangle=\sum_{J,K}\bar{c}_{J}c_{K}^{\prime}\langle
J,K\rangle=\sum_{J,K}\bar{c}_{J}c_{K}^{\prime}\delta_{JK}=\sum_{J}\bar{c}%
_{J}c_{J}^{\prime}=:\mathbf{c}^{\dag}\mathbf{c}^{\prime}%
,\label{eq:sc_prod_HWP}%
\end{equation}
where we have introduced a shorthand notation $\mathbf{c}^{\dag}%
\mathbf{c}^{\prime}$.

\subsection{Commutators of vector operators}

To avoid writing many explicit indices in expressions in the following
sections, let us define a commutator of vector operators and prove several of
its properties. Assuming that $A$ and $B$ are two $D$-dimensional vector
operators, we define a $D\times D$ matrix operator $[  A,B]  $ by%
\begin{equation}
[  A,B]  _{jk}:=[A_{j},B_{k}].\label{eq:comm_vec_A_vec_B}%
\end{equation}
Note that this definition will be more convenient for our purposes than the alternative definition $[A,B]:=A\otimes B^{T}-B\otimes A^{T}$,
i.e., $[A,B]_{jk}:=A_{j}B_{k}-B_{j}A_{k}$. We shall often need the following:

\textbf{Lemma 1}. Let $A$ and $B$ be vector operators, $c$ and $d$ vectors of
numbers, and $C$ and $D$ matrices of numbers. Then%
\begin{align}
\lbrack c^{T}\cdot A,d^{T}\cdot B] &  =c^{T}\cdot\left[  A,B\right]  \cdot
d,\label{eq:comm_vec_A_B_rel_1}\\
\lbrack C\cdot A,D\cdot B] &  =C\cdot\left[  A,B\right]  \cdot D^{T}%
,\label{eq:comm_vec_A_B_rel_2}\\
\lbrack B,A] &  =-[A,B]^{T}.\label{eq:comm_vec_A_B_rel_3}%
\end{align}
\emph{Proof}. Employing Einstein's summation convention over repeated indices,
we have%
\begin{align}
\lbrack c^{T}\cdot A,d^{T}\cdot B] &  =[c_{j}A_{j},d_{k}B_{k}]=c_{j}%
[A_{j},B_{k}]d_{k}=c_{j}\left[  A,B\right]  _{jk}d_{k}%
,\label{eq:comm_vec_ops_1}\\
\lbrack C\cdot A,D\cdot B]_{jk} &  =[(C\cdot A)_{j},(D\cdot B)_{k}%
]=[C_{jl}A_{l},D_{km}B_{m}]\nonumber\\
&  =C_{jl}[A_{l},B_{m}]D_{km}=C_{jl}[A,B]_{lm}D_{mk}^{T}%
,\label{eq:comm_vec_ops_2}\\
\lbrack B,A]_{jk} &  =[B_{j},A_{k}]=-[A_{k},B_{j}]=-[A,B]_{kj}=(-[A,B]^{T}%
)_{jk}.\square\label{eq:comm_vec_ops_3}%
\end{align}
For example, let us re-express the commutators (\ref{eq:comm_ladder}%
)\ between the raising and lowering operators in the matrix form.

\textbf{Proposition 2}. Hagedorn's lowering and raising operators (\ref{eq:A})
and (\ref{eq:Adag}) satisfy the following commutation relations:%
\begin{align}
\left[  A,A\right]   &  =[A^{\dag},A^{\dag}]=0,\\
\lbrack A,A^{\dag}] &  =I_{D}.
\end{align}
\emph{Proof}. Of course, we can obtain these simply by rewriting
(\ref{eq:comm_ladder}) in matrix form. However, let us prove them directly
from the definition of raising and lowering operators [that is, we effectively
also prove (\ref{eq:comm_ladder})]. The relation $\left[  A,A\right]  =0$
follows from the calculation%
\begin{align*}
2\hbar\left[  A,A\right]   &  =-[P^{T}\cdot\hat{x}-Q^{T}\cdot\hat{\xi}%
,P^{T}\cdot\hat{x}-Q^{T}\cdot\hat{\xi}]\\
&  =-[P^{T}\cdot\hat{x},P^{T}\cdot\hat{x}]-[Q^{T}\cdot\hat{\xi},Q^{T}\cdot
\hat{\xi}]+[P^{T}\cdot\hat{x},Q^{T}\cdot\hat{\xi}]+[Q^{T}\cdot\hat{\xi}%
,P^{T}\cdot\hat{x}]\\
&  =-P^{T}\cdot\lbrack\hat{x},\hat{x}]\cdot P-Q^{T}\cdot\lbrack\hat{\xi}%
,\hat{\xi}]\cdot Q+P^{T}\cdot\lbrack\hat{x},\hat{\xi}]\cdot Q+Q^{T}%
\cdot\lbrack\hat{\xi},\hat{x}]\cdot P\\
&  =i\hbar(P^{T}\cdot Q-Q^{T}\cdot P)=0,
\end{align*}
where we have used the definition (\ref{eq:A}) of $A$ in the first step,
bilinearity of the commutator in the second step, relation
(\ref{eq:comm_vec_A_B_rel_2}) in the third step, commutation relations%
\begin{align}
\lbrack\hat{x},\hat{x}] &  =[\hat{q},\hat{q}]=[\hat{\xi},\hat{\xi}]=[\hat
{p},\hat{p}]=0,\label{eq:comm_rel_1}\\
\lbrack\hat{x},\hat{\xi}] &  =[\hat{q},\hat{p}]=i\hbar I_{D}%
\label{eq:comm_rel_2}%
\end{align}
in the fourth step, and the symplecticity conditions (\ref{eq:Q_P_sympl_1}) of
matrices $P$ and $Q$ in the last step. Likewise,%
\begin{align}
2\hbar\left[  A^{\dag},A^{\dag}\right]   &=P^{\dag}\cdot\lbrack\hat{x}%
,\hat{\xi}]\cdot\bar{Q}+Q^{\dag}\cdot\lbrack\hat{\xi},\hat{x}]\cdot\bar
{P}=i\hbar(P^{\dag}\cdot\bar{Q}-Q^{\dag}\cdot\bar{P})=0,\\
2\hbar\left[ A,A^{\dag}\right] &=-P^{T}\cdot\lbrack\hat{x},\hat{\xi}]\cdot\bar
{Q}-Q^{T}\cdot\lbrack\hat{\xi},\hat{x}]\cdot\bar{P}=i\hbar(-P^{T}\cdot\bar
{Q}+Q^{T}\cdot\bar{P})=2\hbar I_{D},
\end{align}
completing the proof.$\square$

\section{Hagedorn wavepackets associated with different Gaussians}

Hagedorn wavepackets with the same Gaussian center are sufficient for finding
expectation values $\langle\hat{O}\rangle:=\langle\psi(t),\hat{O}%
\psi(t)\rangle$ of observables in a state $\psi(t)\equiv h(\mathbf{c}%
_{t},\Lambda_{t})$. If one expresses $\hat{O}\psi(t)$ as another Hagedorn
wavepacket $h(\mathbf{d}_{t},\Lambda_{t})$ with different expansion
coefficients $\mathbf{d}_{t}$ but the same Gaussian center, the expectation
value of the observable is simply obtained as the scalar product $\langle
\hat{O}\rangle=\langle h(\mathbf{c}_{t},\Lambda_{t}),h(\mathbf{d}_{t}%
,\Lambda_{t})\rangle=\mathbf{c}_{t}^{\dag}\mathbf{d}_{t}$ of the two Hagedorn
wavepackets. This procedure is particularly simple if $\hat{O}$ is a
polynomial of position and momentum operators, because then it can be
expressed as a polynomial of Hagedorn's raising and lowering operators and its
action on $h(\mathbf{c}_{t},\Lambda_{t})$ yields another well-defined Hagedorn
wavepacket $h(\mathbf{d}_{t},\Lambda_{t})$ with the same Gaussian center. More
general operators $\hat{O}$ can be expanded in Taylor series about $q_{t}$
and $p_{t}$.

There are situations, however, where one needs to deal with Hagedorn
wavepackets associated with different Gaussians. For example, a wavepacket
spectrum is the Fourier transform of the autocorrelation function%
\begin{equation}
\langle\psi(0)|\psi(t)\rangle=\langle h(\mathbf{c}_{0},\Lambda_{0}%
),h(\mathbf{c}_{t},\Lambda_{t})\rangle,
\end{equation}
which requires the overlap of Hagedorn wavepackets associated with different
Gaussians $\varphi_{0}(\Lambda_{0})$ and $\varphi_{0}(\Lambda_{t})$. In this
section, we shall therefore study Hagedorn operators, functions, and
wavepackets associated with different Gaussians. To simplify notation, we will
use the prime symbol to denote parameters, operators, and multi-indices associated with the
\textquotedblleft second\textquotedblright\ Gaussian, i.e.,
$\Lambda^\prime \equiv (q_t^\prime, p_t^\prime, Q^\prime, P^\prime, S^\prime),
\hat{x}^\prime, \hat{\xi}^\prime, A^\prime, A^{\dagger\prime},K^\prime
$, etc.

\subsection{Commutators of raising and lowering operators}

\textbf{Proposition 3}. Let $A\equiv A(\Lambda)$ and $A^{\prime}\equiv
A(\Lambda^{\prime})$. Then%
\begin{equation}
\lbrack A,A^{\prime\dag}]=\frac{i}{2}\left(  Q^{T}\cdot\bar{P}^{\prime}%
-P^{T}\cdot\bar{Q}^{\prime}\right)  . \label{eq:A_Atildedag_comm}%
\end{equation}
\emph{Proof}. Using definitions (\ref{eq:A}) and (\ref{eq:Adag}) of $A$ and
$A^{\prime\dag}$ gives%
\begin{align}
2\hbar\lbrack A,A^{\prime\dag}]  &  =[P^{T}\cdot\hat{x}-Q^{T}\cdot\hat{\xi
},P^{\prime\dag}\cdot\hat{x}^\prime-Q^{\prime\dag}\cdot\hat{\xi}^\prime]\nonumber\\
&  =-P^{T}\cdot\lbrack\hat{x},\hat{\xi}^\prime]\cdot\bar{Q}^{\prime}-Q^{T}%
\cdot\lbrack\hat{\xi},\hat{x}^\prime]\cdot\bar{P}^{\prime}\\
&  =i\hbar(-P^{T}\cdot\bar{Q}^{\prime}+Q^{T}\cdot\bar{P}^{\prime}),\nonumber
\end{align}
where we have used (\ref{eq:comm_vec_ops_2}) and commutation relations
(\ref{eq:comm_rel_1}) in the second step and commutation relation
(\ref{eq:comm_rel_2})\ in the third step.$\square$

The special case $[A,A^{\dag}]=I_{D}$ for two identical Gaussians follows from
(\ref{eq:A_Atildedag_comm}) and the complex conjugate of the symplecticity
condition (\ref{eq:Q_P_sympl_2}) satisfied by $Q$ and $P$.

\subsection{Bogoliubov transformation}

\textbf{Proposition 4}. Ladder operators associated with different Gaussians
are related by the Bogoliubov transformation%
\begin{align}
A^{\prime} &  =U\cdot A+V\cdot A^{\dag}+v,\label{eq:Atilde_from_A_Adag}\\
A^{\prime\dag} &  =\bar{V}\cdot A+\bar{U}\cdot A^{\dag}+\bar{v}%
,\label{eq:Atildedag_from_A_Adag}%
\end{align}
where matrices $U$, $V$, and vector $v$ are defined in terms of the
Gaussian parameters as%
\begin{align}
U(\Lambda,\Lambda^{\prime}) &  :=\frac{i}{2}\left(  Q^{\prime T}\cdot\bar
{P}-P^{\prime T}\cdot\bar{Q}\right)  ,\label{eq:U_Bogoliubov}\\
V(\Lambda,\Lambda^{\prime}) &  :=\frac{i}{2}\left(  Q^{\prime T}\cdot
P-P^{\prime T}\cdot Q\right)  ,\label{eq:V_Bogoliubov}\\
v(\Lambda,\Lambda^{\prime}) &  :=\frac{i}{\sqrt{2\hbar}}[Q^{\prime T}\cdot(p_t-p^{\prime
}_t)-P^{\prime T}\cdot(q_t-q^{\prime}_t)].\label{eq:v_Bogoliubov}%
\end{align}
\emph{Proof}. Since the definitions (\ref{eq:A}) and (\ref{eq:Adag}) of the
ladder operators hold regardless of the guiding Gaussian, we have%
\begin{align}
A^{\prime} &  =-\frac{i}{\sqrt{2\hbar}}\left(  P^{\prime T}\cdot\hat
{x}^{\prime}-Q^{\prime T}\cdot\hat{\xi}^{\prime}\right)  ,\label{eq:A_prime}\\
A^{\prime\dag} &  =\frac{i}{\sqrt{2\hbar}}\left(  P^{\prime\dag}\cdot\hat
{x}^{\prime}-Q^{\prime\dag}\cdot\hat{\xi}^{\prime}\right)
.\label{eq:Adag_prime}%
\end{align}
The claim of the proposition follows by noting that the displaced position and
momentum operators $\hat{x}^{\prime}$ and $\hat{\xi}^{\prime}$ for the second
Gaussian satisfy
\begin{align}
\hat{x}^{\prime} &  =\hat{q}-q^{\prime}_t=\hat{x}+(q_t-q^{\prime}_t),\\
\hat{\xi}^{\prime} &  =\hat{p}-p^{\prime}_t=\hat{p}+(p_t-p^{\prime}_t),
\end{align}
and by using expressions (\ref{eq:x_from_A_Adag}) and (\ref{eq:xi_from_A_Adag}%
) for $\hat{x}$ and $\hat{\xi}$ in terms of $A$ and $A^{\dag}$.$\square$

Let us make three remarks at this point: (i) The result stated in the
proposition is a generalization of the textbook one-dimensional Bogoliubov
transformation to several degrees of freedom: it includes displacement,
squeezing, and rotation. It is closely related to the multimode squeeze
operators from the quantum optics\ literature~\cite{Ma_Rhodes:1990}. (ii) Note that expressions~(\ref{eq:Atilde_from_A_Adag}%
) and (\ref{eq:Atildedag_from_A_Adag}) for operators $A^{\prime}$ and
$A^{\prime\dag}$ have the desirable property $(A^{\prime})^{\dag}%
=A^{\prime\dag}$. (iii) In the special case $\Lambda^{\prime}=\Lambda$,
equations~(\ref{eq:U_Bogoliubov})--(\ref{eq:v_Bogoliubov}) and symplecticity
conditions~(\ref{eq:Q_P_sympl_1}) and (\ref{eq:Q_P_sympl_2}) for $P$ and $Q$
yield $U\left(  \Lambda,\Lambda\right)  =I_{D}$, $V\left(  \Lambda
,\Lambda\right)  =v(\Lambda,\Lambda)=0$, and therefore $A^{\prime}=A$,
$A^{\prime\dag}=A^{\dag}$, as expected.

\textbf{Corollary 5}. Ladder operators associated with different Gaussians
satisfy the commutation relations%
\begin{align}
\lbrack A,A^{\prime\dag}] &  =U^{\dag},\\
\lbrack A,A^{\prime}] &  =V^{T},\\
\lbrack A^{\dag},A^{\prime\dag}] &  =-V^{\dag},\\
\lbrack A^{\dag},A^{\prime}] &  =-U^{T}.
\end{align}
\emph{Proof}. These follow easily from the transformation rules
(\ref{eq:Atilde_from_A_Adag}) and (\ref{eq:Atildedag_from_A_Adag}):%
\begin{align}
\lbrack A,A^{\prime\dag}] &  =[A,A^{\dag}]\cdot U^{\dag}=U^{\dag
},\label{eq:A_Adag_diff_Gs}\\
\lbrack A,A^{\prime}] &  =[A,A^{\dag}]\cdot V^{T}=V^{T},\label{eq:A_A_diff_Gs}%
\\
\lbrack A^{\dag},A^{\prime\dag}] &  =[A^{\dag},A]\cdot V^{\dag}=-V^{\dag
},\label{eq:Adag_Adag_diff_Gs}\\
\lbrack A^{\dag},A^{\prime}] &  =[A^{\dag},A]\cdot U^{T}=-U^{T}.\square
\label{eq:Adag_A_diff_Gs}%
\end{align}

\subsection{Properties of the transformation matrices $U$ and $V$}

\textbf{Proposition 6}. Transformation matrices $U$ and $V$ have the
properties%
\begin{align}
U\cdot V^{T} &  =V\cdot U^{T},\\
U\cdot U^{\dag}-V\cdot V^{\dag} &  =I_{D}.\label{eq:UUdag-VVdag}%
\end{align}
The first property expresses the symmetry of $U\cdot V^{T}$ and is equivalent
(by complex conjugation) to the relation $\bar{U}\cdot V^{\dag}=\bar{V}\cdot
U^{\dag}$.

\emph{Proof}. Since the operators $A^{\prime}$ and $A^{\prime\dag}$
are defined [see (\ref{eq:A_prime}) and (\ref{eq:Adag_prime})] from
parameters $\Lambda^{\prime}$ in the same way as operators $A$ and $A^{\dag}$
from $\Lambda$, they must satisfy the commutation relations (see Proposition
2) $[A^{\prime},A^{\prime}]=[A^{\prime\dag},A^{\prime\dag}]=0$ and
$[A^{\prime},A^{\prime\dag}]=I_{D}$. Expressing these commutators in terms of
$A$, $A^{\dag}$ using the transformations (\ref{eq:Atilde_from_A_Adag}) and
(\ref{eq:Atildedag_from_A_Adag}) provides the proofs of the properties of $U$
and $V$ matrices:%
\begin{align}
0 &  =[A^{\prime},A^{\prime}]=U\cdot\lbrack A,A^{\dag}]\cdot V^{T}%
+V\cdot\lbrack A^{\dag},A]\cdot U^{T}=U\cdot V^{T}-V\cdot U^{T},\\
I_{D} &  =[A^{\prime},A^{\prime\dag}]=U\cdot\lbrack A,A^{\dag}]\cdot U^{\dag
}+V\cdot\lbrack A^{\dag},A]\cdot V^{\dag}=U\cdot U^{\dag}-V\cdot V^{\dag}.
\end{align}
The commutator $[A^{\prime\dag},A^{\prime\dag}]=0$ does not provide any new
information because it yields the equivalent, complex conjugate of the
property obtained from $[A^{\prime},A^{\prime}]=0$.$\square$

\textbf{Proposition 7}. Transformation matrices $U^{\prime}\equiv
U(\Lambda^{\prime},\Lambda)$ and $V^{\prime}\equiv V(\Lambda^{\prime}%
,\Lambda)$ of the reverse Bogoliubov
transformation are related to the transformation matrices $U\equiv
U(\Lambda,\Lambda^{\prime})$ and $V\equiv V(\Lambda,\Lambda^{\prime})$ of the
forward Bogoliubov transformation by the
equations%
\begin{equation}
U^{\prime}=U^{\dag}\text{ \ \ and \ \ }V^{\prime}=-V^{T}.
\label{eq:U_V_diff_Gs}%
\end{equation}
\emph{Proof}. On one hand, exchanging the roles of forward and reverse
Bogoliubov transformations in (\ref{eq:Adag_A_diff_Gs}) yields
$[A^{\prime\dag},A]=-U^{\prime T}$. On the other hand, applying the general
relation (\ref{eq:comm_vec_A_B_rel_3}) to (\ref{eq:A_Adag_diff_Gs}) gives
$[A^{\prime\dag},A]=-[A,A^{\prime\dag}]^{T}=-(U^{\dag})^{T}$. Equating these
two expressions for $[A^{\prime\dag},A]$ yields (\ref{eq:U_V_diff_Gs}) for
$U^{\prime}$. Likewise, using expression (\ref{eq:comm_vec_A_B_rel_3}) to
relate (\ref{eq:A_A_diff_Gs}) for $[A^{\prime},A]$ to the same equation
for $[A,A^{\prime}]$ , we find that%
\begin{equation}
V^{\prime T}=[A^{\prime},A]=-[A,A^{\prime}]^{T}=(-V^{T})^{T},
\end{equation}
which proves (\ref{eq:U_V_diff_Gs}) for $V^{\prime}$.$\square$

\subsection{Relation to the symplectic structure}

Many quantities discussed above can be expressed more compactly in terms of
the canonical symplectic structure $\omega$ [(\ref{eq:omega_for_z})] on
phase space or its generalization (\ref{eq:omega_for_any_X_Y}). Recalling that $z=\binom{q}{p}$ denotes a phase-space vector and defining a
complex $2D\times D$ matrix%
\begin{equation}
Z=\binom{Q}{P},
\end{equation}
we can use the generalized notation (\ref{eq:omega_for_any_X_Y}) to give
meaning to expressions $\omega(z,z^{\prime})$, $\omega(Z,z)$, $\omega(z,Z)$,
$\omega(Z,Z^{\prime})$. These allow us to express $U,V$, $v$, and $v^{\prime}$
as%
\begin{align}
U(\Lambda,\Lambda^{\prime}) &  =\frac{i}{2}\omega(Z^{\prime},\bar
{Z}),\label{eq:U_from_omega}\\
V(\Lambda,\Lambda^{\prime}) &  =\frac{i}{2}\omega(Z^{\prime}%
,Z),\label{eq:V_from_omega}\\
v(\Lambda,\Lambda^{\prime}) &  =\frac{i}{\sqrt{2\hbar}}\omega(Z^{\prime},z-z^{\prime
}),\label{eq:v_from_omega}\\
v^{\prime}(\Lambda^{\prime},\Lambda) &  =\frac{i}{\sqrt{2\hbar}}\omega(Z,z^{\prime
}-z),\label{eq:v_prime_from_omega}%
\end{align}
and the commutators of raising and lowering operators as%
\begin{align}
\lbrack A,A^{\prime\dag}] &  =U^{\dag}=\frac{i}{2}\omega(Z,\bar{Z}^{\prime
}),\\
\lbrack A,A^{\prime}] &  =V^{T}=-\frac{i}{2}\omega(Z,Z^{\prime}),\\
\lbrack A^{\dag},A^{\prime\dag}] &  =-V^{\dag}=-\frac{i}{2}\omega(\bar{Z}%
,\bar{Z}^{\prime})\\
\lbrack A^{\dag},A^{\prime}] &  =-U^{T}=\frac{i}{2}\omega(\bar{Z},Z^{\prime}).
\end{align}
Note that we also have%
\begin{align}
\omega(Z,Z) &  =0,\\
\omega(Z,\bar{Z}) &  =-2iI_{D}.
\end{align}
Using the symplectic structure, the ladder operators themselves can be written
as%
\begin{align}
A &  =\frac{i}{\sqrt{2\hbar}}\omega(Z,\hat{\zeta}),\\
A^{\dag} &  =-\frac{i}{\sqrt{2\hbar}}\omega(\bar{Z},\hat{\zeta}),
\end{align}
where the operator $\hat{\zeta}$ is defined as
\[
\hat{\zeta}:=\binom{\hat{x}}{\hat{\xi}}%
\]
and, as before, $\hat{x}:=\hat{q}-q_{t}$, $\hat{\xi}:=\hat{p}-p_{t}$ are the
shifted position and momentum operators.

\section{Overlap of Hagedorn functions associated with different Gaussians}

As mentioned above, the autocorrelation function $\left\langle \psi
(0)|\psi(t)\right\rangle $ needed in the evaluation of wavepacket spectra
requires evaluating the overlap
\begin{equation}
\langle\psi,\psi^{\prime}\rangle=\langle h(\mathbf{c},\Lambda),h(\mathbf{c}%
^{\prime},\Lambda^{\prime})\rangle,\label{eq:HWP_sc_pr_diff_G}%
\end{equation}
of Hagedorn wavepackets associated with different Gaussians. This scalar
product could be computed either directly, or indirectly, using the overlap of
Hagedorn functions, as%
\begin{equation}
\langle\psi,\psi^{\prime}\rangle=\sum_{J,K^{\prime}}\bar{c}_{J}\langle
J(\Lambda),K^{\prime}(\Lambda^{\prime})\rangle c_{K^{\prime}}^{\prime
}=\mathbf{c}^{\dag}\mathbf{Mc}^{\prime},\label{eq:HWP_sc_pr_diff_G_via_HF}%
\end{equation}
where
\begin{equation}
M_{JK^{\prime}}:=\langle J(\Lambda),K^{\prime}(\Lambda^{\prime})\rangle
\label{eq:HF_sc_pr_diff_g}%
\end{equation}
is the overlap matrix of Hagedorn functions with different Gaussian centers.

In the direct approach, one could first express the two Hagedorn wavepackets
in the position representation and then evaluate their overlap using various
sophisticated quadrature techniques for highly oscillatory integrals. Instead,
we take the indirect path. Below, we will derive an explicit recursive
expression for the scalar product (\ref{eq:HF_sc_pr_diff_g}), $M_{JK^{\prime}%
}$, in terms of the simple overlap of Gaussians with different parameters,
i.e., in terms of%
\begin{equation}
M_{00^\prime}=\langle g(\Lambda),g(\Lambda^{\prime})\rangle,\label{eq:sc_pr_diff_G}%
\end{equation}
which is well known from the thawed Gaussian wavepacket dynamics. An
analytical expression for this overlap is~\cite{Begusic_Vanicek:2020a}
\begin{equation}
M_{00^\prime}=\frac{(2i)^{D/2}}{\sqrt{\det(Q^{\dagger}\cdot P^{\prime}-P^{\dagger
}\cdot Q^{\prime})}}\exp\left\{  \frac{i}{\hbar}\left[  -\frac{1}{2}\delta
y^{T}\cdot(\delta\Gamma)^{-1}\cdot\delta y+\delta\eta\right]  \right\}
,\label{eq:tg_overlap}%
\end{equation}
where we used the notation $\delta X:=X^{\prime}-\overline{X}$ for matrix,
vector, and scalar tensors
\begin{align}
\Gamma &  :=P\cdot Q^{-1},\\
y &  :=p_{t}-\Gamma\cdot q_{t},\\
\eta &  :=S-\frac{1}{2}\left(  y+p_{t}\right)  ^{T}\cdot q_{t}%
\end{align}
obtained from parameters of each Gaussian. When expression
(\ref{eq:tg_overlap}) is used for evaluating the autocorrelation function at
different times $t$, the branch of the square root in (\ref{eq:tg_overlap})
should be chosen appropriately to ensure continuity of the overlap $M_{00}$ in time.

Next, we derive a system of linear equations satisfied by the overlaps $M_{JK^\prime}$.
The central result of this paper is Proposition 10, in which we solve the system analytically
and thus obtain the promised recurrence relation for these overlaps.

\subsection{System of $2D$ linear equations}

It is useful to group Hagedorn functions $\varphi_{J}$ into \textquotedblleft
shells\textquotedblright\ according to the total excitation $|J|:=J_{1}%
+\cdots+J_{D}$. The $n$th shell is defined to consist of Hagedorn functions
with multi-indices $J$ such that $|J|=n$.

\textbf{Lemma 8}. The overlaps of Hagedorn functions in shell $|J|+1$ with
those in shell $|K^{\prime}|$ and of functions in shell $|J|$ with those in
shell $|K^{\prime}|+1$ satisfy the system%
\begin{align}
\sqrt{J_{j}+1}M_{J+\langle j\rangle,K^{\prime}} &  =\sum_{k=1}^{D}\left(
U_{jk}^{\dag}\sqrt{K_{k}^{\prime}}M_{J,K^{\prime}-\langle k\rangle}-V_{jk}%
^{T}\sqrt{K_{k}^{\prime}+1}M_{J,K^{\prime}+\langle k\rangle}\right)
+v_{j}^{\prime}M_{JK^{\prime}}\label{eq:system_2D_1}\\
\sqrt{K_{k}^{\prime}+1}M_{J,K^{\prime}+\langle k\rangle} &  =\sum_{j=1}%
^{D}\left(  \bar{V}_{kj}\sqrt{J_{j}+1}M_{J+\langle j\rangle,K^{\prime}}%
+\bar{U}_{kj}\sqrt{J_{j}}M_{J-\langle j\rangle,K^{\prime}}\right)  +\bar
{v}_{k}M_{JK^{\prime}}\label{eq:system_2D_2}%
\end{align}
of $2D$ linear equations for $2D$ unknowns $M_{J+\langle j\rangle,K^{\prime}}$
($j=1,\ldots,D$) and $M_{J,K^{\prime}+\langle k\rangle}$ ($k=1,\ldots,D$) in
terms of overlaps $M_{JK^{\prime}}$, $M_{J,K^{\prime}-\langle k\rangle}$, and
$M_{J-\langle j\rangle,K^{\prime}}$ of functions in up to the $|J|$-th and
$|K^{\prime}|$-th shells.

\emph{Proof}. Let us start by evaluating matrix elements of the lowering
operator $A$ associated with the \textquotedblleft bra\textquotedblright%
\ Hagedorn function $\langle J|$ and raising operator $A^{\prime\dag}$ associated with the
\textquotedblleft ket\textquotedblright\ Hagedorn function $|K^\prime \rangle$. On one hand, these matrix
elements are trivially evaluated from the definitions of $A_{j}$ and
$A_{k}^{\prime\dag}$ as%
\begin{align}
\langle J|A_{j}|K^{\prime}\rangle &  =\sqrt{J_{j}+1}\langle J+\langle
j\rangle|K^{\prime}\rangle,\\
\langle J|A_{k}^{\prime\dag}|K^{\prime}\rangle &  =\sqrt{K_{k}^{\prime}%
+1}\langle J|K^{\prime}+\langle k\rangle\rangle.
\end{align}
On the other hand, using the Bogoliubov transformations
(\ref{eq:Atilde_from_A_Adag}) and (\ref{eq:Atildedag_from_A_Adag}),%
\begin{align}
A &  =U^{\prime}\cdot A^{\prime}+V^{\prime}\cdot A^{\prime\dag}+v^{\prime},\\
A^{\prime\dag} &  =\bar{V}\cdot A+\bar{U}\cdot A^{\dag}+\bar{v},
\end{align}
where [see (\ref{eq:U_V_diff_Gs})] $U^{\prime}=U^{\dag}$ \ and $V^{\prime
}=-V^{T}$, we find that%
\begin{align}
\langle J|A_{j}|K^{\prime}\rangle &  =\langle J|U_{jk}^{\prime}A_{k}^{\prime
}+V_{jk}^{\prime}A_{k}^{\prime\dag}+v_{j}^{\prime}|K^{\prime}\rangle
=U_{jk}^{\prime}\langle J|A_{k}^{\prime}|K^{\prime}\rangle+V_{jk}^{\prime
}\langle J|A_{k}^{\prime\dag}|K^{\prime}\rangle+v_{j}^{\prime}\langle
J|K^{\prime}\rangle\nonumber\\
&  =\sum_{k=1}^{D}\left(  U_{jk}^{\prime}\sqrt{K_{k}^{\prime}}\langle
J|K^{\prime}-\langle k\rangle\rangle+V_{jk}^{\prime}\sqrt{K_{k}^{\prime}%
+1}\langle J|K^{\prime}+\langle k\rangle\rangle\right)  +v_{j}^{\prime}\langle
J|K^{\prime}\rangle.
\end{align}
(We have used Einstein's summation convention over repeated indices in the
first but not the second line.) Likewise, for the matrix element of the
raising operator we get%
\begin{align}
\langle J|A_{k}^{\prime\dag}|K^{\prime}\rangle &  =\langle J|\bar{V}_{kj}%
A_{j}+\bar{U}_{kj}A_{j}^{\dag}+\bar{v}_{k}|K^{\prime}\rangle=\bar{V}%
_{kj}\langle J|A_{j}|K^{\prime}\rangle+\bar{U}_{kj}\langle J|A_{j}^{\dag
}|K^{\prime}\rangle+\bar{v}_{k}\langle J|K^{\prime}\rangle\nonumber\\
&  =\sum_{j=1}^{D}\left(  \bar{V}_{kj}\sqrt{J_{j}+1}\langle J+\langle
j\rangle|K^{\prime}\rangle+\bar{U}_{kj}\sqrt{J_{j}}\langle J-\langle
j\rangle|K^{\prime}\rangle\right)  +\bar{v}_{k}\langle J|K^{\prime}\rangle.
\end{align}
Equating the two expressions for $\langle J|A_{j}|K^{\prime}\rangle$ and
repeating the same for $\langle J|A_{k}^{\prime\dag}|K^{\prime}\rangle$ yields
the system (\ref{eq:system_2D_1})--(\ref{eq:system_2D_2}).$\square$

This system can be solved by standard numerical methods, and its sequential
application yields a recursive algorithm for finding all required overlaps
$M_{JK^{\prime}}$: Starting from the zeroth shell $M_{00^{\prime}}$, which is
the overlap (\ref{eq:tg_overlap}) of the two guiding Gaussians, we can
gradually find overlaps of all Hagedorn functions by solving a sequence of
linear systems for additional shells. Next we present a more efficient way
to solve the system.

\subsection{Two systems of $D$ linear equations}

\textbf{Lemma 9}. System~(\ref{eq:system_2D_1})--(\ref{eq:system_2D_2}) of
$2D$ equations is equivalent to two independent systems of $D$ linear
equations for $D$ unknowns. The first system,
\begin{align}
\sqrt{J_{j}+1}M_{J+\langle j\rangle,K^{\prime}} &  =\sum_{k=1}^{D}%
\Big[U_{jk}^{\dag}\sqrt{K_{k}^{\prime}}M_{J,K^{\prime}-\langle k\rangle
}-(V^{T}\cdot\bar{V})_{jk}\sqrt{J_{k}+1}M_{J+\langle k\rangle,K^{\prime}%
}\nonumber\\
&  ~~~~~~~~-(V^{T}\cdot\bar{U})_{jk}\sqrt{J_{k}}M_{J-\langle k\rangle
,K^{\prime}}\Big]+w_{j}M_{JK^{\prime}},~~~j=1,\ldots
,D,\label{eq:system_of_D_eqs_1}%
\end{align}
is for the overlaps $M_{J+\langle j\rangle,K^{\prime}}$, whereas the second
system,%
\begin{align}
\sqrt{K_{k}^{\prime}+1}M_{J,K^{\prime}+\langle k\rangle} &  =\sum_{j=1}%
^{D}\Big[(\bar{V}\cdot U^{\dag})_{kj}\sqrt{K_{j}^{\prime}}M_{J,K^{\prime
}-\langle j\rangle}-(\bar{V}\cdot V^{T})_{kj}\sqrt{K_{j}^{\prime}%
+1}M_{J,K^{\prime}+\langle j\rangle}\nonumber\\
&  ~~~~~~~~+\bar{U}_{kj}\sqrt{J_{j}}M_{J-\langle j\rangle,K^{\prime}%
}\Big]+w_{k}^{\prime}M_{JK^{\prime}},~~~k=1,\ldots
,D,\label{eq:system_of_D_eqs_2}%
\end{align}
is for the overlaps $M_{J,K^{\prime}+\langle k\rangle}$. In
(\ref{eq:system_of_D_eqs_1}) and (\ref{eq:system_of_D_eqs_2}), vectors
$w$ and $w^{\prime}$ are defined as%
\begin{align}
w &  :=-V^{T}\cdot\bar{v}+v^{\prime},\label{eq:w}\\
w^{\prime} &  :=\bar{V}\cdot v^{\prime}+\bar{v}.\label{eq:w_tilde}%
\end{align}
\emph{Proof}. System (\ref{eq:system_2D_1})--(\ref{eq:system_2D_2}) of $2D$
equations is simplified by substituting the former $D$ equations into the
latter $D$ equations and vice versa. This procedure uncouples the equations
for $M_{J+\langle j\rangle,K^{\prime}}$ and $M_{J,K^{\prime}+\langle k\rangle
}$, yielding two independent systems of $D$ equations for $D$ unknowns:
\begin{align}
\sqrt{J_{j}+1}M_{J+\langle j\rangle,K^{\prime}} &  =\sum_{k=1}^{D}%
\Big\{U_{jk}^{\dag}\sqrt{K_{k}^{\prime}}M_{J,K^{\prime}-\langle k\rangle
}-V_{jk}^{T}\Big[\sum_{l=1}^{D}\Big(\bar{V}_{kl}\sqrt{J_{l}+1}M_{J+\langle
l\rangle,K^{\prime}}\nonumber\\
&  ~~~~~+\bar{U}_{kl}\sqrt{J_{l}}M_{J-\langle l\rangle,K^{\prime}}%
\Big)+\bar{v}_{k}M_{JK^{\prime}}\Big]\Big\}+v_{j}^{\prime}M_{JK^{\prime}%
}\label{eq:system_of_D_eqs_1a}\\
\sqrt{K_{k}^{\prime}+1}M_{J,K^{\prime}+\langle k\rangle} &  =\sum_{j=1}%
^{D}\Big\{\bar{V}_{kj}\Big[\sum_{l=1}^{D}\left(  U_{jl}^{\dag}\sqrt
{K_{l}^{\prime}}M_{J,K^{\prime}-\langle l\rangle}-V_{jl}^{T}\sqrt
{K_{l}^{\prime}+1}M_{J,K^{\prime}+\langle l\rangle}\right)  \nonumber\\
&  ~~~~~+v_{j}^{\prime}M_{JK^{\prime}}\Big]+\bar{U}_{kj}\sqrt{J_{j}%
}M_{J-\langle j\rangle,K^{\prime}}\Big\}+\bar{v}_{k}M_{JK^{\prime}%
}\label{eq:system_of_D_eqs_2a}%
\end{align}
If we replace the sum over $k$ in the three terms in square brackets of
(\ref{eq:system_of_D_eqs_1a}) by matrix products and subsequently rename
the dummy index $l$ to $k$, we obtain the system (\ref{eq:system_of_D_eqs_1}).
Repeating this procedure for (\ref{eq:system_of_D_eqs_2a}) yields the
system (\ref{eq:system_of_D_eqs_2}) and completes the proof.$\square$

\subsection{Analytical solution}

\textbf{Proposition 10}. Linear systems~(\ref{eq:system_of_D_eqs_1}) and
(\ref{eq:system_of_D_eqs_2}) have analytical solutions%
\begin{align}
\sqrt{J_{j}+1}M_{J+\langle j\rangle,K^{\prime}}  &  =\sum_{k=1}^{D}\left(
F_{jk}\sqrt{K_{k}^{\prime}}M_{J,K^{\prime}-\langle k\rangle}-G_{jk}\sqrt
{J_{k}}M_{J-\langle k\rangle,K^{\prime}}\right)  +u_{j}M_{JK^{\prime}%
},\label{eq:soln_M_J_plus_j_K}\\
\sqrt{K_{k}^{\prime}+1}M_{J,K^{\prime}+\langle k\rangle}  &  =\sum_{j=1}%
^{D}\left(  G_{kj}^{\prime}\sqrt{K_{j}^{\prime}}M_{J,K^{\prime}-\langle
j\rangle}+F_{kj}^{\prime}\sqrt{J_{j}}M_{J-\langle j\rangle,K^{\prime}}\right)
+u_{k}^{\prime}M_{JK^{\prime}}, \label{eq:soln_M_J_K_plus_k}%
\end{align}
where we have defined matrices%
\begin{align}
W  &  :=(U^{\dag}\cdot U)^{-1},\label{eq:W}\\
W^{\prime}  &  :=(\bar{U}\cdot U^{T})^{-1},\label{eq:W_tilde}\\
F  &  :=W\cdot U^{\dag}\text{ \ \ and \ \ }G:=W\cdot V^{T}\cdot\bar
{U},\label{eq:M_and_N}\\
F^{\prime}  &  :=W^{\prime}\cdot\bar{U}\text{ \ \ and \ \ }G^{\prime
}:=W^{\prime}\cdot\bar{V}\cdot U^{\dag}, \label{eq:M_tilde_and_N_tilde}%
\end{align}
and vectors%
\begin{align}
u  &  :=W\cdot w,\label{eq:z}\\
u^{\prime}  &  :=W^{\prime}\cdot w^{\prime}. \label{eq:z_tilde}%
\end{align}
Note that $W$, $W^{\prime}$, $F$, $F^{\prime}$, $G$, $G^{\prime}$, $u$, and
$u^{\prime}$ are independent of multi-indices $J$ and $K^{\prime}$, and
therefore only depend on the guiding Gaussians.

\emph{Proof}. Moving the middle term in the square brackets in
(\ref{eq:system_of_D_eqs_1}) to the left-hand side yields%
\begin{align}
\sum_{k=1}^{D}\left[  (I_{D}+V^{T}\cdot\bar{V})_{jk}\sqrt{J_{k}+1}M_{J+\langle
k\rangle,K^{\prime}}\right]   &  =\sum_{k=1}^{D}\left[  U_{jk}^{\dag}%
\sqrt{K_{k}^{\prime}}M_{J,K^{\prime}-\langle k\rangle}-(V^{T}\cdot\bar
{U})_{jk}\sqrt{J_{k}}M_{J-\langle k\rangle,K^{\prime}}\right] \nonumber\\
&  ~~~+w_{j}M_{JK^{\prime}}. \label{eq:soln_1st_sys_1}%
\end{align}
The matrix prefactor on the left-hand side can be replaced with $W^{-1}$ since%
\begin{equation}
I_{D}+V^{T}\cdot\bar{V}=I_{D}+V^{\prime}\cdot V^{\prime\dag}=U^{\prime}\cdot
U^{\prime\dag}=U^{\dag}\cdot U=W^{-1},
\end{equation}
where we have used relations $U^{\prime}=U^{\dag}$, $V^{\prime}=-V^{T}$, and
$U\cdot U^{\dag}-V\cdot V^{\dag}=I_{D}$ [see (\ref{eq:U_V_diff_Gs}) and
(\ref{eq:UUdag-VVdag})] and the definition (\ref{eq:W})\ of $W$. Note that $W$
is well-defined since $I_{D}+V^{\prime}\cdot V^{\prime\dag}$ is a
positive-definite and hence invertible Hermitian matrix. This, in turn,
follows because for an arbitrary vector $v\in\mathbb{C}^{D}$%
\begin{equation}
\langle v,(I_{D}+V^{T}\cdot\bar{V})v\rangle=\langle v,v\rangle+\langle
v,V^{T}\cdot\bar{V}\cdot v\rangle=\left\Vert v\right\Vert ^{2}+\left\Vert
\bar{V}\cdot v\right\Vert ^{2}%
\end{equation}
is zero if and only if $v=0$. Multiplying (\ref{eq:soln_1st_sys_1}) from
the left by $W$, we find the explicit solution (\ref{eq:soln_M_J_plus_j_K}).

Likewise, we can move the middle term in the square brackets in
(\ref{eq:system_of_D_eqs_2})\ to the left-hand side and obtain the linear
system%
\begin{align}
\sum_{j=1}^{D}(I_{D}+\bar{V}\cdot V^{T})_{kj}\sqrt{K_{j}^{\prime}%
+1}M_{J,K^{\prime}+\langle j\rangle}  &  =\sum_{j=1}^{D}\left[  (\bar{V}\cdot
U^{\dag})_{kj}\sqrt{K_{j}^{\prime}}M_{J,K^{\prime}-\langle j\rangle}+\bar
{U}_{kj}\sqrt{J_{j}}M_{J-\langle j\rangle,K^{\prime}}\right] \nonumber\\
&  ~~~+w_{k}^{\prime}M_{JK^{\prime}}. \label{eq:soln_2nd_sys_1}%
\end{align}
The matrix prefactor on the left-hand side satisfies
\begin{equation}
I_{D}+\bar{V}\cdot V^{T}=\left(  I_{D}+V\cdot V^{\dag}\right)  ^{T}=(U\cdot
U^{\dag})^{T}=\bar{U}\cdot U^{T}=W^{\prime-1},
\end{equation}
where $W^{\prime}$ is the matrix defined in (\ref{eq:w_tilde}).
Multiplying (\ref{eq:soln_2nd_sys_1}) on the left with $W^{\prime}$, we
obtain the exact solution (\ref{eq:soln_M_J_K_plus_k}).$\square$

Note that all expressions above are explicit since $U$, $V$, $v$, and
$v^{\prime}$ are given by (\ref{eq:U_from_omega}%
)--(\ref{eq:v_prime_from_omega}). In particular, all auxiliary matrices
($U,V,W,F,G,U^{\prime},\ldots$) and vectors ($v,w,u,v^{\prime},\ldots$) depend
only on the parameters $\Lambda \equiv (q,p,Q,P,S)$ and $\Lambda^\prime \equiv(q^{\prime},p^{\prime},Q^{\prime
},P^{\prime},S^{\prime})$ of the two guiding Gaussians. As a result, these
auxiliary matrices and vectors, which appear repeatedly in the recursive
expressions, do not have to be recomputed for different overlaps
$M_{JK^\prime}$ as long as $\Lambda$ and $\Lambda^\prime$ do not change.
In the Appendix, we provide
explicit, nonrecursive expressions for the first and second shells in general
dimensions and describe how the recursive expressions simplify in one dimension.

Ideally, one should come up also with a direct recursive algorithm for
converting the scalar product (\ref{eq:HWP_sc_pr_diff_G}) of Hagedorn
wavepackets directly to the scalar product (\ref{eq:sc_pr_diff_G}) of
Gaussians. Note, however, that computing the overlap matrix $\mathbf{M}$ first
allows a quick calculation of overlap of any Hagedorn wavepackets associated
to the same two Gaussians. In contrast, the direct algorithm would be specific
for the given two Hagedorn wavepackets.

\section{Numerical experiments}

To verify the analytical expressions (\ref{eq:soln_M_J_plus_j_K}) and (\ref{eq:soln_M_J_K_plus_k}) for the overlaps of arbitrary Hagedorn
functions, we performed several numerical experiments.

\subsection{Implementation and numerical details}

The recursive algebraic expressions described in the previous section were
implemented in Python with the NumPy package~\cite{software_numpy:2020}. For
numerical integration, we used the default quadrature integration procedure
(\texttt{nquad}) included in the SciPy package~\cite{software_scipy:2020},
which in turn calls subroutines from the Fortran library
QUADPACK~\cite{software_quadpack:1983}. For simplicity, we set $\hbar=1$ in the
numerical experiments.

\subsection{Comparison with numerical integration results}

To verify the correctness and assess the accuracy of our algorithm, we
compared its results to numerically evaluated overlaps $\langle J|K^{\prime
}\rangle$ of basis functions from a pair of two-dimensional Hagedorn bases
associated with two different Gaussian wavepackets, $g[\Lambda] \equiv \varphi_{0}[\Lambda]$ and
$g^{\prime}[\Lambda^{\prime}] \equiv \varphi_{0}^{\prime}[\Lambda^{\prime}]$, with
\[
\begin{aligned} q &= (0.033, 0.141),& {q^{\prime}} &= (-0.943, -0.657), \\ p &= (0.371, 0.668),& {p^{\prime}} &= (-0.386, 0.787), \\ S &= 0,& {S^{\prime}} &= -0.62,\\ Q &= \begin{pmatrix} \phantom{-}1.626 & \,-0.256 \\ -0.256 & \,\phantom{-}1.409 \end{pmatrix},& P &=\begin{pmatrix} -0.009+0.633i & \,\phantom{+}0.051+0.115i \\ \phantom{+}0.059+0.115i & \,-0.009+0.731i \end{pmatrix}\\ Q^{\prime} &= \begin{pmatrix} \phantom{+}2.042 & \,-0.235 \\ -0.235 & \,\phantom{+}1.268 \end{pmatrix},& P^{\prime} &=\begin{pmatrix} -0.004+0.500i & \,\phantom{+}0.022+0.093i \\ \phantom{+}0.035+0.093i & \,-0.004+0.806i \end{pmatrix}. \end{aligned}
\]
The overlap integrals $\langle{J}|{K}^{\prime}\rangle$ for $\max J_{j},\max
K_{k}^{\prime}\leq2$ (for a total of 81 pairs of basis functions) were
calculated with both our algebraic approach and numerical integration.

For all 81 integrals considered, the absolute differences between the
algebraic and numerical results, for both real and imaginary parts, were
smaller than $10^{-10}$. Table \ref{tab:num_diff} shows the overlaps computed
using the algebraic approach for nine selected pairs of basis functions and
the differences from the numerical results (the numerical results themselves
were omitted from the Table due to the tiny differences between algebraic and
numerical results). Full results are available in the Supplementary material.
Analogous comparisons are carried out for four other pairs of Hagedorn bases
with randomly generated parameters, and the differences between the algebraic
algorithm and numerical integration results were of a similar order of
magnitude (see the Supplementary material). These results reassure us that our
algebraic scheme as well as its Python implementation were correct. \newcolumntype{R}{>{$}r<{$}}

\begin{table}[htb]
\caption{\label{tab:num_diff}Algebraic results for the overlaps between selected pairs of Hagedorn basis functions $\langle {J} | K^{\prime} \rangle$ and differences (algebraic results minus numerical ones) from numerical results. The full comparison table of algebraic and numerical results (up to 15 decimal places) for all 81 pairs are available in the Supplementary material.}
\begin{indented}
\item[]
\begin{tabular}{ccRR}
\br
$J$ & $K^{\prime}$ &\multicolumn{1}{r}{\quad Overlap (algebraic)} &  \multicolumn{1}{r}{\quad Alg.\,$-$\,Num. ($\times 10^{-11}$)\quad}   \\ \mr
(0, 0) & (0, 0) & 0.47376-0.08503i & -0.0006\phantom{}+0.001\phantom{}i \\
(0, 0) & (2, 1) & -0.06856-0.18401i & -0.009\phantom{0}+0.07\phantom{0}i \\
(0, 2) & (1, 0) & -0.06029-0.06231i & \phantom{+}0.0007\phantom{}+0.001\phantom{}i \\
(1, 0) & (1, 2) & -0.01413-0.02169i & \phantom{+}0.2\phantom{000}+0.2\phantom{00}i \\
(1, 1) & (0, 2) & 0.02424-0.33445i & \phantom{+}0.01\phantom{00}-0.03\phantom{0}i \\
(1, 1) & (1, 1) & -0.04884-0.10501i & -0.04\phantom{00}+0.09\phantom{0}i \\
(2, 0) & (1, 2) & 0.03729-0.08187i & \phantom{+}0.3\phantom{000}-1\phantom{.000}i \\
(2, 1) & (1, 1) & -0.17283-0.20042i & \phantom{+}0.2\phantom{000}-0.6\phantom{00}i \\
(2, 1) & (2, 2) & -0.10699-0.15887i & -2\phantom{.0000}+5\phantom{.000}i \\
\br
\end{tabular}\end{indented}
\end{table}

We also note that, despite being implemented in an interpreted (hence
relatively slow) language, the algebraic algorithm in Python was much faster
than the numerical computation: on the same computer, the computation of all 81 overlaps took on
average 0.1 seconds using the algebraic algorithm, while the numerical
integrals took about one minute. Improvements may be possible using a more
efficient implementation (e.g., in Fortran) or more advanced numerical
schemes, but the time of computation using our algebraic method is in any case
satisfactory for applications in chemical dynamics.

\subsection{Approximation of wavefunctions in another Hagedorn basis}

As a secondary check and a small demonstration of the properties of Hagedorn
bases, we test the self-consistency of our scheme using the property of the
Hagedorn functions as a complete orthonormal basis of $L^{2}(\mathbb{R}^{D})$.

Given a wavepacket $\psi$ expanded in a Hagedorn basis associated with the
Gaussian $\varphi_{0}[\Lambda]$, we approximate it by projecting the
wavepacket onto another Hagedorn basis associated with a different Gaussian
${\varphi}_{0}^{\prime}[{\Lambda}^{\prime}]$:
\begin{equation}
\psi=\sum_{J}c_{J}\varphi_{J}\simeq\sum_{|K^{\prime}|\leq K_{\text{max}%
}^{\prime}}{c}_{K^{\prime}}^{\prime}{\varphi}_{K^{\prime}}^{\prime}=\sum
_{K}\langle{\varphi}_{K^{\prime}}^{\prime}|\psi\rangle{\varphi}_{K^{\prime}%
}^{\prime}={\psi}_{K_{\text{max}}^{\prime}}^{\prime}.\label{eq:approx_psi}%
\end{equation}
For a given dimensionality $D$, two sets of Gaussian parameters were randomly
generated. The wavepacket $\psi$ is defined as the linear combination of four
basis functions $\varphi_{J}$ (with $|J| := \sum_{i=1}^{D} J_{i} < 5$) with
the same weight ($c_{J} = 0.5$). The approximate wavepacket ${\psi}^{\prime}$
was computed following (\ref{eq:approx_psi}) with a ``simplex" basis set
defined by the requirement that
all multi-indices $K^{\prime}$ satisfy $|K^{\prime}|
\leq K^{\prime}_{\text{max}}$. The overlap integral $\langle\psi|{\psi
}^{\prime}_{K^{\prime}_{\text{max}}} \rangle$ was then calculated for increasing values of $K^{\prime
}_{\text{max}}$.

\begin{table}[!hbt]
\caption{\label{tab:approx}Overlaps $|\langle \psi|{\psi}^{\prime}_{K^{\prime}_{\text{max}}} \rangle|$ (rounded to four significant digits) between the original wavefunction $\psi$ and its projection ${\psi}^{\prime}_{K^{\prime}_{\text{max}}}$ onto another
Hagedorn basis with a restricted number of basis functions.}
\begin{indented}
\item[]\ \\
(a) $D=3$\\
\begin{tabular}{rrD{.}{.}{4,4}}\br
\rule{0pt}{2.5ex} $K^{\prime}_{\text{max}}$ &\quad$\#$ of basis functions & \multicolumn{1}{c}{$\quad|\langle \psi|{\psi}^{\prime}_{K^{\prime}_{\text{max}}} \rangle|$\quad} \\\mr
0& 1                       &\quad 0.04688                        \\
2& 10                       &\quad 0.1719\phantom{0}                      \\
4& 35                         & \quad 0.4525\phantom{0}                           \\
8& 165                         &\quad   0.7988\phantom{0}                           \\
16& 969                          &\quad  0.9893\phantom{0}                            \\
32& 6\,545                         &\quad  1.000\phantom{0}                         \\\br
\end{tabular}
\\
\item[]
(b) $D=5$\\
\begin{tabular}{rrD{.}{.}{4,4}}\br
\rule{0pt}{2.5ex} $K^{\prime}_{\text{max}}$ &\quad$\#$ of basis functions & \multicolumn{1}{c}{$\quad|\langle \psi|{\psi}^{\prime}_{K^{\prime}_{\text{max}}} \rangle|$\quad} \\\mr
0& 1                       &\quad 0.006848  \quad                      \\
2& 21                       &\quad 0.04019\phantom{0}   \quad                      \\
4& 126                         & \quad 0.1318\phantom{00} \quad                          \\
8& 1\,287                          &\quad   0.4565\phantom{00} \quad                         \\
16& 20\,349                          &\quad  0.9203\phantom{00}  \quad                         \\
32& 435\,897                          &\quad  0.9999\phantom{00} \quad                       \\\br
\end{tabular}
\end{indented}
\end{table}

Table \ref{tab:approx} presents the results for two examples, one in three
dimensions and another in five dimensions, with parameters specified in the
Supplementary material. We observe that the overlap between the
wavepacket and its projection clearly converges towards unity as the number of basis
functions increases. These results demonstrate that our algorithm is
consistent with the algebraic structure and properties of Hagedorn bases.

\subsection{Propagated wavepacket: comparison with the split-operator Fourier
method}

Hagedorn wavepackets, like the thawed Gaussian wavepacket, are exact solutions
of the TDSE with a harmonic potential. In a harmonic system, the coefficients
of the Hagedorn basis functions remain unchanged while the Gaussian parameters
evolve with classical-like equations of motion~\cite{Lasser_Lubich:2020}. Here,
we used a three-dimensional harmonic potential to propagate a Hagedorn
wavepacket $\psi(t)$ and calculate the autocorrelation function $\langle
\psi(0)|\psi(t)\rangle$ along the trajectory. The same simulation was carried
out with the split-operator Fourier
method~\cite{Feit_Steiger:1982,book_Tannor:2007,Kosloff_Kosloff:1983,Roulet_Vanicek:2019}. Since
the wavepacket was continuously displaced, squeezed, and rotated under the
influence of the potential, the comparison with the numerical quantum
benchmark effectively verifies our expressions and implementation for many
different Hagedorn bases.

\begin{figure}[!tbh]
\centering
\includegraphics{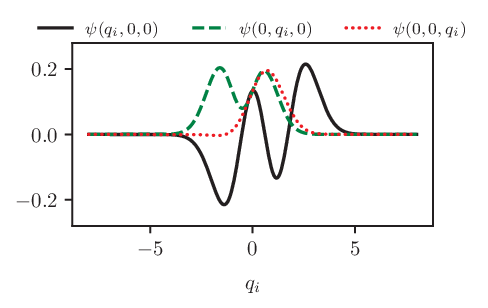} \caption{Select cross-sections of the
three-dimensional initial wavepacket $\psi(q_1, q_2, q_3)$.}
\label{fig:3d_wavef}
\end{figure}

The initial wavepacket, with a unit mass, was chosen to be the linear
combination of the $(3,0,0)$ and $(1,2,1)$ basis functions with equal weights
(see Figure \ref{fig:3d_wavef} for cross-sections of the initial
wavefunction). The associated Gaussian parameters corresponded to the ground
state of another harmonic potential that is displaced, squeezed and rotated
compared to the potential used for propagation (the parameters are available
in the Supplementary material). The wavepacket was propagated for 2000 steps
with a time step of 0.1 and the autocorrelation function was computed every
five steps.

\begin{figure}[!htb]
\centering
\includegraphics{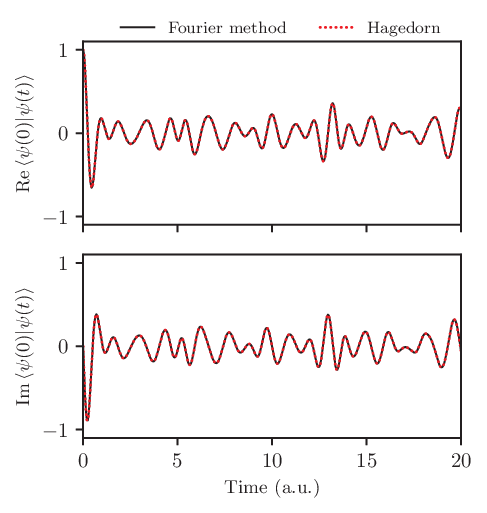} \caption{Comparison of the autocorrelation functions
obtained with the split-operator Fourier method and Hagedorn approach.}
\label{fig:3d_ct_spec}
\end{figure}

Figure \ref{fig:3d_ct_spec} shows that the autocorrelation function computed
with the Hagedorn approach agrees perfectly with the autocorrelation function
obtained with the split-operator Fourier method. In this example, the Fourier
method required $32^{3}$ grid points to obtain a converged result, whereas the
gridless Hagedorn wavepacket dynamics only needed the propagation of the five
Gaussian parameters by solving a system of first-order ordinary differential
equations. The Hagedorn approach can easily treat both the propagation and the
computation of overlap integrals in much higher dimensions than the grid-based
Fourier method whose cost grows exponentially with the number of degrees of freedoms.

\section{Conclusions}

We have discussed properties of Hagedorn functions and wavepackets associated
with two different Gaussians. In particular, we have derived algebraic
recurrence expressions for the overlap between two Hagedorn functions with
different Gaussian centers and numerically demonstrated that both our
expressions and their implementation are correct, efficient, and robust.

With these expressions available, Hagedorn wavepackets should find more
applications in spectroscopy, particularly in situations where a non-Gaussian
initial state is generated (e.g., in single vibronic level
fluorescence~\cite{Tapavicza:2019} or Herzberg--Teller
spectroscopy~\cite{Patoz_Vanicek:2018}) or where anharmonicity results in the
occupation of excited Hagedorn functions.

\section*{Acknowledgements}

The authors acknowledge the financial support from the European Research
Council (ERC) under the European Union's Horizon 2020 Research and Innovation
Programme (Grant Agreement No.~683069--MOLEQULE) and thank Lipeng Chen for useful discussions.

\appendix

\section{Special cases}

The algebraic recursive expressions (\ref{eq:soln_M_J_plus_j_K}%
)--(\ref{eq:soln_M_J_K_plus_k}) for the overlaps are valid for any dimension
$D$ and any excitation shells\ numbered by the total excitation $|J|$ and
$|K^{\prime}|$. Here we describe how the general expressions simplify
substantially if one is only interested in arbitrary excitations of
one-dimensional systems or in low excitations of arbitrary-dimensional systems.

\subsection{One-dimensional case}

For $D=1$, the multi-indices become ordinary indices and the solutions
(\ref{eq:soln_M_J_plus_j_K})--(\ref{eq:soln_M_J_K_plus_k}) are:%
\begin{align}
M_{J+1,K^{\prime}}  &  =\left[  \sqrt{J+1}|U|^{2}\right]  ^{-1}(\bar{U}%
\sqrt{K^{\prime}}M_{J,K^{\prime}-1}-V\bar{U}\sqrt{J}M_{J-1,K^{\prime}%
}+wM_{JK^{\prime}}),\\
M_{J,K^{\prime}+1}  &  =\left[  \sqrt{K^{\prime}+1}|U|^{2}\right]  ^{-1}%
(\bar{V}\bar{U}\sqrt{K^{\prime}}M_{J,K^{\prime}-1}+\bar{U}\sqrt{J}%
M_{J-1,K^{\prime}}+w^{\prime}M_{JK^{\prime}}).
\end{align}

\subsection{First shell in many dimensions}

Let us evaluate the overlaps between the zeroth and first shells for any $D$. We set
$J:=0$ and $K^{\prime}:=0$ in the general solutions
(\ref{eq:soln_M_J_plus_j_K}) for $M_{J+\langle j\rangle,K^{\prime}}$ and
(\ref{eq:soln_M_J_K_plus_k}) for $M_{J,K^{\prime}+\langle k\rangle}$ to find%
\begin{align}
M_{\langle j\rangle,0}  &  =u_{j}M_{00},\\
M_{0,\langle k\rangle}  &  =u_{k}^{\prime}M_{00},
\end{align}
where the scalar quantity%
\begin{equation}
M_{00}:=M_{|J|=0,|K^{\prime}|=0}=\langle0|0^{\prime}\rangle
\end{equation}
is the overlap of the guiding Gaussians. In matrix form, the solution can be
written as
\begin{align}
M_{10}  &  =uM_{00},\\
M_{01}  &  =u^{\prime}M_{00},
\end{align}
where%
\begin{equation}
M_{10}:=M_{|J|=1,|K^{\prime}|=0}\text{ \ \ and \ \ }M_{01}%
:=M_{|J|=0,|K^{\prime}|=1}\text{ }%
\end{equation}
are the $D$-vectors of overlaps between the zeroth and first shells.

To find overlaps of the first shells, we set $J:=0$ and $K^{\prime
}:=\left\langle k\right\rangle $ in the general solution
(\ref{eq:soln_M_J_plus_j_K}) for $M_{J+\langle j\rangle,K^{\prime}}$ and get%
\begin{equation}
M_{\langle j\rangle,\langle k\rangle}=F_{jk}M_{00}+u_{j}M_{0,\langle k\rangle
},
\end{equation}
which, in matrix notation, becomes%
\begin{align}
M_{11}  &  =FM_{00}+u\otimes M_{01}^{T}\nonumber\\
&  =(F+u\otimes u^{\prime T})M_{00}%
\end{align}
where
\begin{equation}
M_{11}:=M_{|J|=1,|K^{\prime}|=1}%
\end{equation}
is a $D\times D$ overlap matrix of the first shell functions. If $U$ itself is
invertible, so is $U^{\dag}$ since $(U^{\dag})^{-1}=(U^{-1})^{\dag}$. As a
result, $W=U^{-1}\cdot(U^{\dag})^{-1}$, $F=U^{-1}$, and%
\begin{equation}
M_{11}\overset{U\text{ invertible}}{=}(U^{-1}+u\otimes u^{\prime T})M_{00}.
\end{equation}

\subsection{Second shell in many dimensions}

To evaluate the overlap matrix $M_{20}$, let us set $J:=\langle j\rangle$,
$j:=k$, $K^{\prime}:=0$, and $k:=l$ in the general expression
(\ref{eq:soln_M_J_plus_j_K}) for $M_{J+\langle j\rangle,K^{\prime}}$ and find
that%
\begin{equation}
\sqrt{\delta_{jk}+1}M_{\langle j\rangle+\langle k\rangle,0}=(-GM_{00}+u\otimes
M_{10}^{T})_{jk}=(-G+u\otimes u^{T})_{jk}M_{00}.
\end{equation}
To compute the $3$-tensor $M_{21}$, we set $J:=\langle j\rangle$, $j:=k$,
$K^{\prime}:=\langle l\rangle$, $k:=m$ in the expression
(\ref{eq:soln_M_J_plus_j_K}) for $M_{J+\langle j\rangle,K^{\prime}}$ and find
that%
\begin{align}
\sqrt{\delta_{jk}+1}M_{\langle j\rangle+\langle k\rangle,\langle l\rangle}  &
=F_{kl}M_{\langle j\rangle,0}-G_{kj}M_{0,\langle l\rangle}+u_{k}M_{\langle
j\rangle\langle l\rangle}\nonumber\\
&  =[F_{kl}u_{j}-G_{kj}u_{l}^{\prime}+u_{k}(F+u\otimes u^{\prime T}%
)_{jl}]M_{00}.
\end{align}
Note that the left-hand sides of equations for $M_{20}$ and $M_{21}$ are
obviously symmetric w.r.t. exchange of $j$ and $k$. It may be numerically
advantageous to take the symmetric average of the right-hand side of the
corresponding equations.

To find $M_{02}$, let us set $J:=0$ and $K^{\prime}:=\langle j\rangle$ in the
expression for $M_{J,K^{\prime}+\langle k\rangle}$:%
\begin{equation}
\sqrt{\delta_{jk}+1}M_{0,\langle j\rangle+\langle k\rangle}=(G^{\prime}%
M_{00}+u^{\prime}\otimes M_{01}^{T})_{kj}=(G^{\prime}+u^{\prime}\otimes
u^{\prime T})_{kj}M_{00}.
\end{equation}
To find $M_{12}$, we set $J:=\langle j\rangle$, $K^{\prime}:=\langle k\rangle
$, $k:=l$, $j:=m$ in the general expression for $M_{J,K^{\prime}+\langle
k\rangle}$ and obtain%
\begin{align}
\sqrt{\delta_{kl}+1}M_{\langle j\rangle,\langle k\rangle+\langle l\rangle}  &
=G_{lk}^{\prime}M_{\langle j\rangle,0}+F_{lj}^{\prime}M_{0,\langle k\rangle
}+u_{l}^{\prime}M_{\langle j\rangle\langle k\rangle}\nonumber\\
&  =[G_{lk}^{\prime}u_{j}+F_{lj}^{\prime}u_{k}^{\prime}+u_{l}^{\prime
}(F+u\otimes u^{\prime T})_{jk}]M_{00}.
\end{align}
Finally, to find $M_{22}$ from $M_{12}$, let us set $J:=\langle j\rangle$,
$K^{\prime}:=\langle k\rangle+\langle m\rangle$, $k:=r$, $j:=l$ in the general
expression for $M_{J+\langle j\rangle,K^{\prime}}$:%
\begin{align}
&  \sqrt{\delta_{jl}+1}M_{\langle j\rangle+\langle l\rangle,\langle
k\rangle+\langle m\rangle}\nonumber\\
&  =\sum_{r=1}^{D}\left(  F_{lr}\sqrt{\delta_{kr}+\delta_{mr}}M_{\langle
j\rangle,\langle k\rangle+\langle m\rangle-\langle r\rangle}-G_{lr}%
\sqrt{\delta_{jr}}M_{\langle j\rangle-\langle r\rangle,\langle k\rangle
+\langle m\rangle}\right)  +u_{l}M_{\langle j\rangle,\langle k\rangle+\langle
m\rangle}\nonumber\\
&  =\sum_{r=1}^{D}\left(  F_{lr}\sqrt{\delta_{kr}+\delta_{mr}}M_{\langle
j\rangle,\langle k\rangle+\langle m\rangle-\langle r\rangle}\right)
-G_{lj}M_{0,\langle k\rangle+\langle m\rangle}+u_{l}M_{\langle j\rangle
,\langle k\rangle+\langle m\rangle}%
\end{align}
It is convenient to distinguish cases $k=m$ and $k\neq m$:%
\begin{align}
&  \sqrt{\delta_{jl}+1}M_{\langle j\rangle+\langle l\rangle,\langle
k\rangle+\langle m\rangle}\overset{k\neq m}{=}F_{lk}M_{\langle j\rangle
,\langle m\rangle}+F_{lm}M_{\langle j\rangle,\langle k\rangle}-G_{lj}%
M_{0,\langle k\rangle+\langle m\rangle}+u_{l}M_{\langle j\rangle,\langle
k\rangle+\langle m\rangle},\\
&  \sqrt{\delta_{jl}+1}M_{\langle j\rangle+\langle l\rangle,2\langle k\rangle
}\overset{k=m}{=}F_{lk}\sqrt{2}M_{\langle j\rangle,\langle k\rangle}%
-G_{lj}M_{0,2\langle k\rangle}+u_{l}M_{\langle j\rangle,2\langle k\rangle}.
\end{align}

\section*{Data availability statement}
All data that support the findings of this study are included within the article (and any supplementary files).

\section*{References}
\bibliographystyle{iopart-num}
\bibliography{hagedorn_overlap_v33_iop}

\end{document}